\documentclass[twocolumn,a4paper]{article}
\usepackage[utf8]{inputenc}
\usepackage[T1]{fontenc}
\usepackage{amsmath,amssymb,amsfonts}
\usepackage{graphicx}
\usepackage{xcolor}
\usepackage{hyperref}
\usepackage{cite}
\usepackage{geometry}
\usepackage{cuted}
\usepackage{graphicx}
\usepackage{array}
\usepackage{multirow, booktabs, xcolor}

\title{\vspace{-1.5em}\textbf{Geometry-Induced Termination of the Repetitive Penrose Process in Rotating Simpson–Visser Black Holes}}
\author{\textbf{Mohammad Ali S. Afshar}\textsuperscript{1,2,3}\thanks{m.a.s.afshar@gmail.com}, \textbf{Mohammad Reza Alipour}\textsuperscript{1,2,3}\thanks{mohamad.alipour.1994@gmail.com},\\ [0.3em]\textbf{Saeed Noori Gashti}\textsuperscript{2,3}\thanks{sn.gashti@du.ac.ir; saeed.noorigashti70@gmail.com} and \textbf{J. Sadeghi}\textsuperscript{1}\thanks{pouriya@ipm.ir}. \\[0.5em]
\small\textsuperscript{1}Department of Theoretical Physics, Faculty of Basic Sciences, University of Mazandaran, P. O. Box 47416-95447, Babolsar, Iran \\
\small\textsuperscript{2}School of Physics, Damghan University, P. O. Box 3671641167, Damghan, Iran \\
\small\textsuperscript{3}Center for Theoretical Physics, Khazar University, 41 Mehseti Street, Baku, AZ1096, Azerbaijan}
\date{}
\newcommand{\keywords}[1]{\par\vspace{0.3em}\noindent\textbf{Keywords:} #1}
\begin{document}

\maketitle
\thispagestyle{empty}
\begin{strip}
\vspace{-3cm}

\begin{abstract}

In this work, we examine the Repetitive Penrose Process (RPP) in the rotating Simpson-Visser spacetime (R-S-V), whose parameter space includes regular black holes, black-bounce geometries, and traversable wormholes.

Assuming that the regularization parameter remains unchanged throughout the evolution, the analysis shows that the endpoint of the repetitive process is not always determined solely by the conventional minimum-spin condition. Instead, for sufficiently large values of the regularization parameter, the evolving solution may leave the parameter domain corresponding to the original two horizon black hole branch before the dynamical spin limit is reached. Within the present framework, this provides an additional geometry-induced condition that limits the continuation of the iterative sequence.

The numerical results further show that the relative importance of the dynamical and geometry-induced termination mechanisms depends sensitively on the regularization parameter. For small deformations, the evolution remains qualitatively similar to that of the Kerr spacetime. As the regularization parameter increases, however, the cumulative extracted energy, the number of admissible Penrose iterations, and the efficiency of the process are progressively reduced. We also examine how the extracted energy, the final irreducible mass, and two complementary efficiency measures vary with both the regularization parameter and the particle decay radius.

Overall, the present analysis indicates that, within the R-S-V geometry, the underlying spacetime structure influences not only the cumulative efficiency of the repetitive Penrose process but also the parameter range over which the iterative evolution remains self-consistent. These results highlight the role that spacetime geometry can play in shaping the long-term evolution of idealized Penrose-type energy extraction processes in regular rotating black-hole spacetimes.

\end{abstract}

\end{strip}

\newpage 
\section{Introduction}
The emergence of the black hole concept, and in particular the discovery of the Kerr solution, marked a major milestone in our understanding of the final stages of gravitational collapse in massive stars. Beyond their role as compact objects predicted by general relativity, rotating black holes possess a remarkable property: part of their rotational energy can, in principle, be extracted through physical processes operating in the ergosphere or the surrounding magnetosphere.

Rotating black holes therefore provide a unique environment in which strong-field gravity, electromagnetic processes, and relativistic plasma physics intersect. According to the Christodoulou mass formula for Kerr black holes~\cite{1}, the total mass can be decomposed into an irreducible component associated with the horizon area and a rotational component that is, in principle, extractable. For an extremal Kerr black hole, the maximum extractable rotational energy is approximately $29\%$ of the total mass-energy~\cite{2}. Although this fraction may appear modest, the corresponding amount of energy is enormous for astrophysical black holes. For example, for a supermassive black hole with mass $M\sim10^{9}M_{\odot}$, the available rotational energy is of the order of $10^{74}$~eV, placing rotating black holes among the most energetic systems predicted by classical general relativity.

The possibility of extracting this rotational energy has motivated extensive theoretical research over the past several decades. These efforts have led to a variety of proposed mechanisms that rely on different aspects of relativistic gravity, electromagnetic fields, and plasma dynamics. Among these mechanisms, the Penrose process, first proposed by Roger Penrose in 1969~\cite{3}, provided the first explicit demonstration that rotational energy can be extracted from a rotating black hole.

The Penrose process is based on the existence of negative-energy trajectories, defined with respect to an observer at infinity, within the ergoregion of a Kerr black hole. In this region, frame dragging prevents the existence of stationary observers and allows particles with appropriate angular momentum to occupy negative-energy states. In the original scenario, an infalling particle entering the ergosphere splits into two fragments. One fragment is captured by the black hole after entering a negative-energy trajectory, thereby reducing the black hole's mass and angular momentum, while the second fragment escapes to infinity carrying more energy than the incident particle. As a result, part of the black hole's rotational energy is transferred to the escaping fragment.

Despite its conceptual importance, the classical Penrose process faces significant challenges with regard to its astrophysical viability. For an extremal Kerr black hole, the maximum theoretical efficiency is:
\[
\eta_{\rm PP}=\frac{\sqrt{2}-1}{2}\approx20.7\% .
\]
More importantly, the mechanism requires the decay fragments to acquire a relative velocity exceeding approximately half the speed of light in order for one of them to reach a negative-energy trajectory. Such an instantaneous acceleration of neutral particles is generally considered unlikely under realistic astrophysical conditions. Consequently, although the Penrose process established the theoretical possibility of extracting rotational energy, considerable effort has been devoted to developing alternative mechanisms that may operate under realistic astrophysical conditions.
The limitations of the original Penrose process motivated the development of alternative energy extraction mechanisms that based  on electromagnetic fields, plasma dynamics, or collective interactions rather than discrete particle decay. Over the past five decades, this field has gradually evolved from idealized particle-based models toward increasingly realistic magnetohydrodynamic descriptions of black hole environments.

\paragraph{The Blandford--Znajek mechanism.}
Among the proposed mechanisms, the Blandford--Znajek (BZ) process~\cite{4,5} is widely regarded as the standard theoretical framework for the electromagnetic extraction of rotational energy from astrophysical black holes. In this picture, a rotating black hole is threaded by magnetic field lines sustained by electric currents in the surrounding accretion flow. Frame dragging twists these field lines, generating an electromotive force that drives large-scale electric currents and carries energy away from the black hole primarily through an outward Poynting flux.

The BZ mechanism is widely regarded as one of the leading candidates for powering the relativistic jets observed in active galactic nuclei and possibly other energetic compact-object systems. Its operation, however, requires a sufficiently magnetized, nearly force-free magnetosphere supported by an electron-positron plasma. Consequently, its efficiency depends not only on the black-hole spin but also on the global magnetospheric structure and the magnetic flux threading the event horizon.
\paragraph{The Magnetic Penrose Process.}
A different approach to overcoming the limitations of the original Penrose process was proposed in the form of the Magnetic Penrose Process (MPP)~\cite{5,6}. In this mechanism, charged particles interact with an external electromagnetic field as they move through the ergosphere. The electromagnetic contribution to the conserved energy enables particles to reach negative-energy states without requiring unrealistically large relative velocities.

Unlike the original Penrose process, whose efficiency is limited to approximately $21\%$, the MPP exhibits different efficiency regimes depending on the magnetic-field strength, black-hole spin, and particle charge. In sufficiently magnetized environments, the extraction efficiency may formally exceed $100\%$, reflecting the fact that part of the energy carried by the escaping particle is supplied by the rotational energy of the black hole rather than by the incident particle alone~\cite{6}. Nevertheless, despite its attractive theoretical properties, the MPP still relies on discrete particle interactions and therefore does not directly describe the collective plasma dynamics expected in realistic accretion flows.
\paragraph{The Bañados--Silk--West effect.}

Another influential development was proposed by Bañados, Silk, and West (BSW) ~\cite{7}, who demonstrated that the center-of-mass energy of two particles colliding arbitrarily close to the horizon of an extremal Kerr black hole can, under suitable conditions, become arbitrarily large if one of the particles possesses a critical angular momentum.

The BSW effect is not itself an energy extraction mechanism. Rather, it provides a framework in which extremely energetic particle collisions may occur in the strong gravitational field surrounding rapidly rotating black holes. For this reason, it has often been discussed in connection with high-energy particle production and as a possible precursor to secondary energy extraction processes such as the collisional Penrose process.
\paragraph{The Comisso--Asenjo mechanism.}
More recently, Comisso and Asenjo~\cite{8} proposed an energy-extraction mechanism based on relativistic magnetic reconnection within the ergosphere. In this scenario, rapid reconnection converts magnetic energy into plasma kinetic energy, producing oppositely directed outflows. Under suitable conditions, the decelerated outflow acquires negative energy with respect to infinity and is captured by the black hole, whereas the accelerated outflow escapes carrying away rotational energy.

This mechanism combines relativistic magnetic reconnection with the Penrose concept of negative-energy states. Analytical studies suggest that, for rapidly rotating black holes embedded in strongly magnetized plasmas, magnetic reconnection can provide an efficient channel for extracting rotational energy and may generate intermittent high-energy emission associated with reconnection events. Under suitable conditions, the extracted power may become comparable to, or even exceed, that predicted by the Blandford-Znajek mechanism.
\paragraph{The Repetitive Penrose Process}
Although the original Penrose process and its various extensions have been studied extensively as isolated energy-extraction events, a natural question is how a sequence of such events evolves over time. More specifically, one may ask whether repeated Penrose-type interactions can progressively extract the rotational energy of a Kerr black hole and, if so, what determines the ultimate limit of this process.

Early analyses, most notably those of Misner, Thorne, and Wheeler~\cite{9,10}, suggested that an idealized sequence of Penrose events could, in principle, continue until the black hole approached the Schwarzschild limit through the gradual loss of angular momentum. However, this conclusion relied on a simplified treatment in which the changes induced by each extraction event were incorporated only approximately.

Subsequent studies, particularly those by Ruffini and collaborators~\cite{10,11}, showed that a fully self-consistent treatment leads to a markedly different picture. As successive Penrose events occur, the black-hole mass and angular momentum decrease, while the irreducible mass increases in accordance with Hawking's area theorem. Consequently, only a fraction of the initial rotational energy is recovered as extracted energy, while a substantial portion contributes irreversibly to the growth of the horizon area. 

The repetitive Penrose process therefore terminates before the black hole reaches the non-rotating Schwarzschild state. The endpoint is instead determined by the kinematic conditions required for negative-energy trajectories to exist inside the ergosphere. Once these conditions are no longer satisfied, further Penrose events become impossible and the extraction process terminates ~\cite{10,11}.

This behaviour highlights an important feature of repetitive rotational energy extraction. Although each individual Penrose event can, in principle, extract energy from the black hole, the cumulative evolution is intrinsically nonlinear because each extraction event alters the black-hole parameters that determine the subsequent evolution. As a result, the overall efficiency of the repetitive process is significantly lower than the maximum rotational energy predicted by the Christodoulou decomposition.
More recent investigations have further explored repetitive Penrose processes in generalized rotating black hole geometries, indicating that additional gravitational or electromagnetic charges may impose even stronger restrictions on the cumulative extraction efficiency ~\cite{12,13,14,15,16}. Recent studies of the repetitive Penrose process in Kerr--Newman spacetimes have shown that, for charged particles interacting with the background electromagnetic field, the cumulative extraction efficiency can differ substantially from that of the neutral case under appropriate parameter choices~\cite{17}. These results suggest that the endpoint of repeated energy extraction is highly sensitive to the underlying spacetime geometry.\\
Despite the remarkable progress achieved by macroscopic plasma-based energy extraction mechanisms, the repetitive Penrose process remains an attractive theoretical framework for investigating the fundamental limits imposed solely by spacetime geometry. Unlike magnetohydrodynamic mechanisms, whose efficiency depends on complex plasma physics and magnetic field configurations, the repetitive Penrose process is governed primarily by the geodesic structure of the underlying spacetime and therefore provides a cleaner probe of the geometric constraints on rotational energy extraction.
For this reason, we adopt the RPP as the theoretical framework of the present work. Its well-defined termination point offers a natural setting for investigating how modifications of the spacetime geometry influence the long-term evolution of energy extraction.
In this work we consider the R-S-V spacetime ~\cite{18,19}, a particularly interesting family of geometries whose physical interpretation depends sensitively on the values of its parameters. Different regions of the parameter space describe qualitatively distinct compact objects, ranging from Kerr-like regular black holes to black-bounce geometries and traversable wormholes. This rich phenomenology makes the model an ideal laboratory for investigating how changes in the spacetime geometry influence rotational energy extraction.
The central question addressed in this work is whether the cumulative back-reaction associated with repetitive Penrose-type energy extraction can drive the R-S-V spacetime across different regions of its parameter space. More specifically, we investigate whether the secular evolution of the black hole parameters induced by repeated energy extraction can alter the qualitative nature of the underlying geometry, and how such transitions affect both the efficiency and the ultimate endpoint of the extraction process.

\section{Methodology: The Nonlinear Repetitive Penrose Process}
To model the cumulative extraction of rotational energy, we employ the nonlinear RPP framework. Unlike early linear formulations, which treated the background spacetime as unchanged between successive decay events, the present approach updates the macroscopic black-hole parameters after each extraction step in a self-consistent manner while preserving global energy conservation. 

Consider an incident particle (particle 0) with rest mass $\mu_0$ falling from infinity into the ergoregion, where it splits into two fragments: a captured fragment (particle 1) that crosses the event horizon, and an escaping fragment (particle 2) that reaches asymptotic infinity. The decay kinematics are determined by local four-momentum conservation at the splitting point. Denoting the dimensionless specific energy, axial angular momentum, and radial momentum of the $i$-th particle as $\hat{E}_i = E_i/\mu_i$, $\hat{p}_{\phi i} = p_{\phi i}/(\mu_i M)$, and $\hat{p}_{ri} = p_{ri}/\mu_i$, respectively, and defining the mass ratios $\tilde{\mu}_i = \mu_i/\mu_0$, the conservation laws read:
\begin{align}
    \hat{E}_0 &= \tilde{\mu}_1 \hat{E}_1 + \tilde{\mu}_2 \hat{E}_2, \\
    \hat{p}_{\phi 0} &= \tilde{\mu}_1 \hat{p}_{\phi 1} + \tilde{\mu}_2 \hat{p}_{\phi 2}, \\
    \hat{p}_{r0} &= \tilde{\mu}_1 \hat{p}_{r1} + \tilde{\mu}_2 \hat{p}_{r2}.
\end{align}

For equatorial motion in a stationary, axisymmetric spacetime, the mass-shell condition $g^{\mu\nu}p_\mu p_\nu = -\mu_i^2$ dictates the radial dynamics. Solving this constraint for the radial momentum yields:
\begin{equation}
    \hat{p}_{ri}^2 = -\frac{g^{tt}}{g^{rr}} (\hat{E}_i - \hat{V}^+_i)(\hat{E}_i - \hat{V}^-_i),
\end{equation}
where the effective potentials $\hat{V}^\pm_i$ are defined as ~\cite{11}:
\begin{equation}
    \hat{V}^\pm_i = \frac{-g^{t\phi}M \hat{p}_{\phi i} \pm \sqrt{(g^{t\phi}M \hat{p}_{\phi i})^2 - g^{tt}(g^{\phi\phi}M^2 \hat{p}_{\phi i}^2 + 1)}}{g^{tt}}.
\end{equation}
Physical trajectories must be future-directed and timelike, implying $dt/d\tau > 0$. This requirement excludes the lower branch $\hat{V}^-_i$, restricting the physical domain to $\hat{E}_i \ge \hat{V}^+_i$. 

Maximum energy extraction is achieved when the captured particle occupies the lowest negative-energy state permitted by the spacetime geometry, which corresponds to its classical turning point, $\hat{E}_1 = \hat{V}^+_1$. Under this assumption, conservation of radial momentum implies $\hat{p}_{r0} = \hat{p}_{r2}$. Suppose instead that this common radial momentum is non-zero. substituting this result into the mass-shell condition together with the energy and angular-momentum conservation laws leads to $\hat{E}_1 = (g^{t\phi}/g^{tt})M \hat{p}_{\phi 1}$. However, evaluating the coordinate time derivative $dt/d\tau$ for this specific energy yields exactly zero, which violates the strict timelike geodesic requirement. 

The assumption of non-zero radial momenta must therefore be discarded. The physically admissible configuration that maximizes the extracted energy while preserving the timelike nature of the trajectories is the \textit{triple turning-point condition}, wherein the radial momenta of all three particles vanish simultaneously at the decay radius $\hat{r}_d$ ~\cite{11}:
\begin{equation}\label{6}
\begin{split}
   & \hat{p}_{r0} = \hat{p}_{r1} = \hat{p}_{r2} = 0 \quad \implies\\& \quad \hat{E}_i = \hat{V}^+_i \quad \text{for } i \in \{0, 1, 2\}.\\
    \end{split}
\end{equation}

\subsection{Analytic Solution and Nonlinear Iterative Evolution}
Under the triple turning-point condition, and taking the incident energy $\hat{E}_0$, the captured angular momentum $\hat{p}_{\phi 1}$, and the mass ratio $\nu=\mu_2/\mu_1$ as free input parameters, the conservation equations admit a closed-form analytic solution. The angular momentum of the incident particle and the energy of the captured fragment are then obtained as~\cite{11}:
\begin{align}
    \hat{p}_{\phi 0} &= \frac{g^{t\phi} \hat{E}_0 + \sqrt{(g^{t\phi} \hat{E}_0)^2 - g^{\phi\phi}(1 + g^{tt} \hat{E}_0^2)}}{M g^{\phi\phi}}, \\
    \hat{E}_1 &= \frac{g^{t\phi} \hat{p}_{\phi 1} M - \sqrt{(g^{t\phi} \hat{p}_{\phi 1} M)^2 - g^{tt}(g^{\phi\phi} \hat{p}_{\phi 1}^2 M^2 + 1)}}{g^{tt}}.
\end{align}
The mass ratio of the captured fragment follows from the corresponding quadratic constraint and is given by:
\begin{equation}
    \tilde{\mu}_1 = \frac{\mathcal{A} + \sqrt{\mathcal{D}}}{\mathcal{B}},
\end{equation}
where the coefficients $\mathcal{A}$, $\mathcal{B}$, and the discriminant $\mathcal{D}$ are algebraic functions of the metric components evaluated at the decay radius, the black-hole parameters, and the prescribed kinematic variables (their explicit expressions are given in Appendix). The conserved quantities of the escaping fragment are then obtained from:
\begin{equation}
    \hat{E}_2 = \frac{\hat{E}_0 - \tilde{\mu}_1 \hat{E}_1}{\nu}, \quad \hat{p}_{\phi 2} = \frac{\hat{p}_{\phi 0} - \tilde{\mu}_1 \hat{p}_{\phi 1}}{\nu}.
\end{equation}

Following the absorption of the negative-energy fragment, the black hole's mass $M$ and angular momentum $L$ are updated accordingly. If the parameters prior to the $n$-th decay are $(M_{n-1}, L_{n-1})$, the post-decay parameters become:
\begin{align}
    M_n &= M_{n-1} + \mu_0 \tilde{\mu}_{1, n-1} \hat{E}_{1, n-1}, \\
    L_n &= L_{n-1} + M_{n-1} \mu_{1, n-1} \hat{p}_{\phi 1}.
\end{align}
This induces a nonlinear evolution in the dimensionless spin parameter $\hat{a}_n = L_n/M_n^2$, governed by:
\begin{equation}
    \Delta \hat{a}_{n-1} = \frac{L_n}{M_n^2} - \frac{L_{n-1}}{M_{n-1}^2}.
\end{equation}
For more general rotating black-hole solutions, additional spacetime parameters may also require iterative updating. For example, depending on the underlying geometry, quantities such as the cosmological constant $\Lambda$, the electric charge $Q$, or the dimensionless Gauss--Bonnet coupling $\hat{\alpha}$ may evolve according to the corresponding conservation laws adopted for the model under consideration.

\subsection{Thermodynamic Constraints and Efficiency Metrics}
The thermodynamic limit of the extraction process is determined by the irreducible mass $M_{\mathrm{irr}}$, which is proportional to the square root of the event-horizon area. The extractable energy reservoir is defined as $E_{\mathrm{extractable}}=M_{\mathrm{local}}-M_{\mathrm{irr}}$, where $M_{\mathrm{local}}$ denotes the effective mass entering the corresponding black-hole solution. For the Kerr spacetime, $M_{\mathrm{local}}=M$, whereas in more general geometries additional contributions may arise depending on the specific definition of the conserved mass adopted for the background.

To quantify the performance of the iterative process, we introduce two complementary efficiency measures. The Energy Return on Investment (EROI), denoted by $\zeta_n$, measures the net energy gain relative to the total energy injected by the incident particles ~\cite{17}:
\begin{equation}
    \zeta_n = \frac{E_{\text{extracted}, n}}{n E_0},
\end{equation}
where $E_{\text{extracted}, n} = M_0 - M_n$ is the cumulative extracted energy after $n$ iterations. 

The Energy Utilization Efficiency (EUE), denoted by $\Xi_n$, measures the fraction of the depleted extractable reservoir that is successfully harvested as escaping radiation, rather than contributing irreversibly to the increase of the horizon area ~\cite{17}:
\begin{equation}
    \Xi_n = \frac{E_{\text{extracted}, n}}{E_{\text{extractable}, 0} - E_{\text{extractable}, n}}.
\end{equation}
In the idealized reversible limit (the Christodoulou-Ruffini limit), the irreducible mass remains constant ($\Delta M_{\text{irr}} = 0$), yielding $\Xi_n \to 1$. In realistic repetitive sequences, $\Xi_n$ remains substantially below unity due to the irreversible area increase mandated by Hawking's area theorem.

\subsection{Termination Conditions}
The repetitive decay sequence is necessarily finite. The termination of the process is determined by a set of strict kinematic and thermodynamic constraints that must be satisfied simultaneously at every iteration step $n$:
\begin{enumerate}
    \item \textbf{Kinematic Feasibility:} The decay must exhibit a positive mass defect, i.e. $\mu_0>\mu_1+\mu_2$, ensuring the rest mass of the incident particle exceeds the combined rest masses of the fragments ($\mu_0 - \mu_1 - \mu_2 > 0$).
    \item \textbf{Negative Energy Infall:} The captured fragment must possess negative energy as measured by an asymptotic observer ($\hat{E}_1 < 0$).
    \item \textbf{Thermodynamic Consistency:} The extractable energy must remain strictly positive ($E_{\text{extractable}, n} > 0$), and the irreducible mass must not decrease ($\Delta M_{\text{irr}} \ge 0$).
    \item \textbf{Marginal Stability (Effective Potential Constraints):} For the trajectories to remain physically valid, the classical turning points of the incident particle (0) and the escaping fragment (2) must lie on the outer side of the maximum of their respective effective potentials $\hat{V}^+_i$. Conversely, the turning point of the captured fragment (1) must lie on the inner side. The iterative sequence terminates when one of these turning points reaches the corresponding maximum of the effective potential, beyond which the required orbit no longer exists. Mathematically, this critical marginal stability condition is expressed as:
    \begin{equation}
    \begin{split}
        &\hat{V}^+_i(\hat{r}_d) = \hat{E}_i, \quad \text{and}\\ & \quad \left. \frac{d \hat{V}^+_i}{d \hat{r}} \right|_{\hat{r}=\hat{r}_d} = 0, \quad \text{for } i \in \{0, 1, 2\}.\\
        \end{split}
    \end{equation}
\end{enumerate}
For each particle, solving the marginal stability condition yields a critical minimum spin threshold, $\hat{a}_{\text{min}, i}$, required for the corresponding orbit to exist. The repetitive Penrose process inevitably terminates at the iteration step where the updated spin parameter drops below the most restrictive of these bounds:
\begin{equation}
    \hat{a}_n < \max \left( \hat{a}_{\text{min}, 0}, \hat{a}_{\text{min}, 1}, \hat{a}_{\text{min}, 2} \right).
\end{equation}
Depending on the underlying black-hole solution and its additional parameters or couplings, the hierarchy of these minimum spin thresholds may shift, altering which particle ultimately governs the termination of the repetitive process.
\section{The Model: Geometric Structure and Physical Properties of the Rotating Simpson-Visser(R-S-V) Black Hole}
One of the longstanding challenges in classical general relativity is the appearance of curvature singularities in black-hole interiors. Among the proposed resolutions, the R-S-V geometry stands out due to its minimalist and mathematically elegant approach. In its static form \cite{18,19}, the standard radial coordinate $r$ is replaced by $\sqrt{r^2+\xi^2}$. In this construction, the parameter $\xi$ plays the role of a regularization length scale that determines the radius of a minimal two-dimensional sphere at the core of the geometry. In this picture, timelike trajectories reach a minimum areal radius, $r_{\rm min}=\xi$, at the bounce surface before extending smoothly into another asymptotic region (or another region of the same spacetime), rather than terminating at a curvature singularity. This nonsingular transition is commonly referred to as a \emph{black bounce}

Since astrophysical black-hole candidates are expected to possess angular momentum, extending the geometry to the rotating case is essential for astrophysical applications. The R-S-V metric, obtained through the Newman-Janis algorithm and representing a specific subclass of Johannsen's parametrized non-Kerr geometries \cite{20}, is expressed in Boyer-Lindquist coordinates $(t, r, \theta, \phi)$ (with $G=c=1$) as:
\begin{equation}\label{1}
\begin{split}
&ds^2 =\\ & -\left(1 - \frac{2M\sqrt{r^2 + \xi^2}}{\rho^2}\right) dt^2 + \frac{\rho^2}{\Delta} dr^2 +\\ & \rho^2 d\theta^2 - \frac{4aM\sqrt{r^2 + \xi^2} \sin^2\theta}{\rho^2} dt d\phi + \frac{\Sigma \sin^2\theta}{\rho^2} d\phi^2,\\
\end{split}    
\end{equation}
where the metric functions are defined as:
\begin{align}
    \rho^2 &= r^2 + \xi^2 + a^2 \cos^2\theta, \\
    \Delta &= r^2 + \xi^2 + a^2 - 2M\sqrt{r^2 + \xi^2}, \\
    \Sigma &= (r^2 + \xi^2 + a^2)^2 - \Delta a^2 \sin^2\theta.
\end{align}
Here, $M$ denotes the ADM mass, $a$ is the spin parameter (specific angular momentum), and $\xi$ is the regularization parameter characterizing deviations from the Kerr geometry. In the limit $\xi\rightarrow0$, the metric reduces to the Kerr solution, while the simultaneous limits $\xi\rightarrow0$ and $a\rightarrow0$ recover the Schwarzschild spacetime.

The causal structure and event horizons are determined by the roots of the equation $\Delta(r) = 0$. The interplay between the spin parameter $a$ and the regularization parameter $\xi$ divides the parameter space into three distinct regimes, as summarized in Table~\ref{tab:horizon_classification}.

   \begin{table*}[th]\label{tab:horizon_classification}
	\centering
	\caption{Classification of R-S-V spacetime according to the parameter $\xi$ \cite{19}.}
	\label{tab:horizon_classification}
	\begin{tabular}{c c c}
		\hline
		Range of $\xi$ & Horizons & Spacetime \\
		\hline
		$[0,M-\sqrt{M^2-a^2})$ 
		& $r_-,\,r_+$ 
		& Regular black hole \\
		
		$[M-\sqrt{M^2-a^2}  , M+\sqrt{M^2-a^2})$ 
		& $r_+$
		& Single-horizon regular black hole \\
		
		$[M+\sqrt{M^2-a^2},\infty)$ 
		& No horizon 
		& Wormhole \\
		\hline
	\end{tabular}
\end{table*}
In this work, we restrict our analysis to the two-horizon regular black-hole regime,
\[
0\le\xi<M-\sqrt{M^2-a^2},
\]
for which both the outer and inner horizons are present. In this domain, the spacetime possesses both an outer ($r_+$) and an inner ($r_-$) event horizon. Within this regime, increasing $\xi$ shifts the outer horizon inward, so that $r_+$ becomes progressively smaller than that of a Kerr black hole with the same mass and spin.

Before introducing our analysis, we briefly summarize several physical properties of the R-S-V geometry that are relevant to particle dynamics and observational studies:

\begin{itemize}
    \item \textbf{Invariance of ISCO Quantities and Radiative Efficiency:}\\ A noteworthy feature of the geodesic structure is that, although the parameter $\xi$ alters the radial coordinate of the innermost stable circular orbit ($r_{\text{ISCO}}$) and modifies the radial profiles of the conserved specific energy ($E$), angular momentum ($L$), and angular velocity ($\Omega$), the exact numerical values of $E$, $L$, and $\Omega$ evaluated precisely at $r_{\text{ISCO}}$ remain unchanged as $\xi$ varies. As a consequence, the radiative efficiency ($\eta \approx 1 - E_{\text{ISCO}}$) of the thin accretion disk is identical to that of a Kerr black hole with the same spin, showing that, within this model, spacetime regularization does not modify the global radiative efficiency of a thin accretion disk.\cite{19}.
    \item \textbf{Suppression of Local Radiative Properties:} In contrast to the global efficiency, the local radiative characteristics of the accretion disk are modified. An increase in the regularization parameter $\xi$ leads to a suppression of the peak values of the radiative flux ($F$) and the effective temperature ($T_{\text{eff}}$), while slightly shifting the radial position of these maxima inward. This reduction effect becomes more pronounced for black holes with higher spin parameters \cite{19}.
    \item \textbf{Photon Ring Broadening and Redshift/Blueshift Signatures:}\\ In ray-tracing simulations, while the primary shadow remains largely unchanged, increasing $\xi$ leads to a measurable broadening of the photon ring width and the higher-order lensed images. Furthermore, at high observer inclination angles (e.g., $60^\circ$ or $80^\circ$), the parameter $\xi$ enhances the extent of the blueshifted regions on the approaching side of the disk, thereby increasing the overall intensity contrast in the observed image. These effects may provide additional observational signatures that help distinguish this geometry from the Kerr spacetime \cite{19}.
    \item \textbf{Exotic Field Sources and Energy Conditions:}\\ The removal of the central singularity requires non-standard matter sources. The R-S-V geometry cannot be sourced by a canonical scalar field or standard Maxwell electrodynamics alone. It requires an exact solution to the Einstein field equations sourced by a combination of a minimally coupled phantom scalar field (with negative kinetic energy) and a magnetic field within the framework of nonlinear electrodynamics (NED). Within this construction, this matter content is required to violate the Null Energy Condition (NEC) in the vicinity of the bounce throat, a prerequisite for maintaining the regular core or traversable wormhole structure \cite{21}.
        \item \textbf{Kerr Mimicry and Shadow Degeneracy:}\\ In the asymptotic exterior, the metric approaches the Kerr geometry. Within the two-horizon regular black hole regime, the bounce region remains hidden behind the photon sphere. Since the shadow boundary is governed by the photon capture region, which is insensitive to these internal modifications,the size and shape of the shadow remain very similar with those of a Kerr black hole of identical mass and spin. This motivates the exploration of complementary observables capable of distinguishing the two geometries.\cite{19,20}.
\end{itemize}
 \section{RPP In the R-S-V Black Hole: Competition Between Dynamical and Geometry-Induced Termination Mechanisms}
A distinctive feature of the RPP in the R-S-V spacetime arises from the interplay between the secular evolution of the black-hole mass and angular momentum and the fixed regularization parameter $\xi$, which determines the geometrical class of the spacetime throughout the evolution. 
Unlike the mass and angular momentum,  the parameter $\xi$ is assumed to remain unchanged during the evolution. Consequently, each Penrose event modifies only the dynamical parameters of the black hole, while the regularization scale continues to characterize the same underlying geometry. Although this assumption is simple, it has important consequences once the Penrose process is applied repeatedly.
In every iteration, the extraction of energy changes both the black-hole mass and its dimensionless spin parameter. Since the admissible range of $\xi$ depends on these quantities, the region of parameter space corresponding to the initial two-horizon configuration evolves together with the black-hole parameters. Consequently, the black-hole solution traces a trajectory through the parameter space instead of remaining at a fixed location.

This behaviour has no analogue in the Kerr spacetime, where no additional geometric parameter accompanies the evolution. Besides the familiar dynamical termination of the iterative Penrose process associated with the minimum allowed spin, the evolution may also encounter the boundary of the parameter domain defining the original two-horizon branch. If this boundary is crossed, the spacetime no longer satisfies the assumptions under which the iterative sequence was initially constructed.

This additional limitation does not originate from the kinematics of particle splitting or from the energetics of the Penrose process itself. Instead, it is a direct consequence of the geometrical structure of the R-S-V solution space. It is therefore convenient to distinguish two independent mechanisms governing the endpoint of the iterative Penrose process.

The first mechanism is purely dynamical and has already been discussed in previous sections. As rotational energy is continuously extracted, the dimensionless spin decreases until the minimum value compatible with further Penrose events is reached. At this point, no additional extraction of rotational energy is possible, and the iterative sequence terminates in the conventional sense.

The second mechanism has a different origin. During the iterative evolution, the changing mass-spin configuration may eventually leave the admissible parameter domain corresponding to the original two-horizon black-hole solution while the spin still exceeds its minimum allowed value. In this situation, the assumptions underlying the initial two-horizon solution are no longer satisfied, and the iterative sequence cannot be consistently continued within the same geometrical branch.

This behaviour introduces an additional termination mechanism for the RPP, which we term the \textit{structural termination condition}. The iterative process is no longer governed solely by the kinematic spin threshold ($\hat{a}_n < \hat{a}_{\text{min}}$), but by a competition between a dynamical constraint associated with the extraction of rotational energy and a geometry-induced constraint associated with the admissible parameter domain of the underlying spacetime. The spacetime geometry therefore acts not only as the background on which the Penrose process occurs, but also as a restrictive constraint on the continuation of the iterative evolution.

The central objective of the following analysis is to determine which of these two mechanisms sets the ultimate endpoint of the RPP across the parameter space of the R-S-V black hole.

A geometry-induced limitation of this type is not unique to the present model. In our previous investigation of the iterative Penrose process in the four-dimensional Einstein-Gauss-Bonnet black-hole spacetime \cite{22}, we found that the Gauss-Bonnet coupling parameter could also restrict the continuation of the iterative sequence for appropriate choices of the model parameters. Nevertheless, the physical origin of this restriction differs fundamentally from that encountered in the R-S-V geometry. In the Einstein-Gauss-Bonnet case, also the coupling parameter  modifies the spacetime geometry and consequently alters the kinematical conditions required for successive Penrose events. For sufficiently large values of the coupling parameter, the iterative process  reaches a stage at which no further energy extraction satisfying the Penrose conditions is possible. Throughout this evolution, the black-hole solution remains within the same geometrical branch.

Unlike the Einstein--Gauss--Bonnet case \cite{22}, where the iterative sequence terminates because the Penrose process itself can no longer proceed, the R-S-V spacetime exhibits a qualitatively different termination mechanism. Here, the iterative sequence terminates because the evolving black-hole solution leaves the geometrical branch on which the iterative framework is based. The former corresponds to the exhaustion of the physical conditions required for rotational energy extraction, whereas the latter results from the loss of compatibility between the evolving black-hole parameters and the geometrical branch defining the initial spacetime.
\begin{figure}
    \centering
    \includegraphics[width=\linewidth]{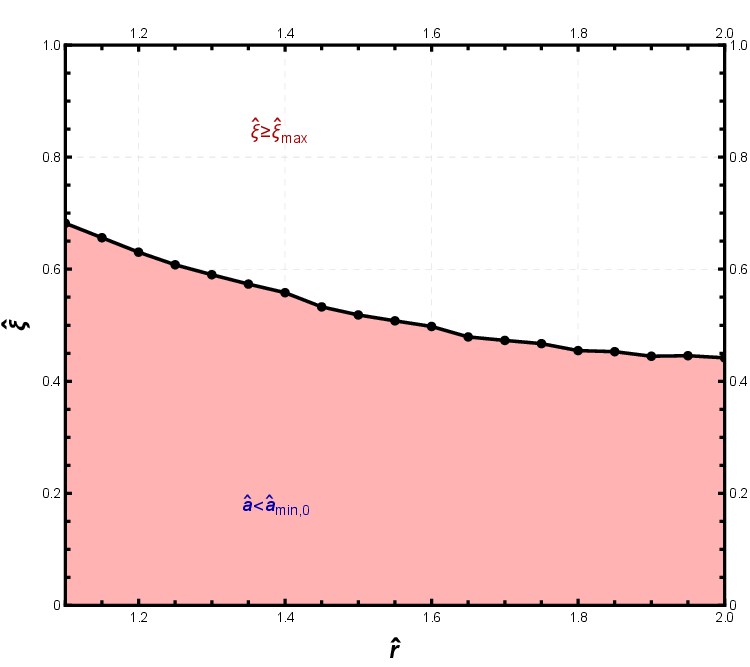}
    \caption{The parameter space of the RPP in the $(\hat{r}_d, \hat{\xi})$ plane. The black solid curve with dots marks the boundary separating two termination regimes. The shaded pink region below the curve indicates the domain where the process is terminated by the minimum spin condition ($\hat{a} < \hat{a}_{\text{min},0}$). The white region above the curve corresponds to the domain where the process is terminated by the structural condition ($\hat{\xi} \ge \hat{\xi}_{\text{max}}$), i.e., the black hole transitions out of the two-horizon regime.}
    \label{1}
\end{figure}

The numerical results summarized in Fig.~\ref{1} and Tables~2 and~3 (in the Appendix) show that the endpoint of the RPP is governed by two distinct termination regimes, with the transition between them controlled primarily by the regularization parameter $\hat{\xi}$. This transition constitutes the central result of the present work.
 
In table 2, for sufficiently small values of $\xi$, or equivalently its dimensionless counterpart $\hat{\xi}=\xi/M$, the evolution closely resembles that found in the Kerr spacetime. Throughout this stage, the evolving solution remains within the admissible parameter domain of the two-horizon R-S-V black hole, and the geometry-induced constraint therefore remains inactive. 

Unless stated otherwise, all numerical calculations presented in this work employ the representative parameter set $\hat{E}_0=1$, $\hat{p}_{\phi1}=-19.434$, $\nu=\mu_2/\mu_1=0.78345$, $\mu_0=10^{-2}M_0$, and $M_0=M=1$. For this parameter set, the above behaviour persists up to approximately $\hat{\xi}\simeq0.44$, beyond which a transition to the geometry-induced termination regime is observed.
\begin{table*}[p]
\centering
\setlength{\arrayrulewidth}{0.5pt}
\renewcommand{\arraystretch}{1.2}
\scriptsize
\setlength{\tabcolsep}{3pt}
\resizebox{\textwidth}{!}{%
\begin{tabular}{cccccccccccccc}
\toprule
 $\xi_0$ & $n$ & $M/M_0$ & $\hat{a}_n$ & $\hat{\xi}_n$ & $E_{\text{extractable}}/M_0$ & $E_{\text{extracted}}/M_0$ & $M_\text{irr}/M_0$ & $\tilde{\mu}_{1,n}$ & $\hat{E}_{1,n}$ & $\hat{p}_{\phi0,n}$ & $\hat{a}_{\min,0,n}$ & $\zeta_n$ & $\Xi_n$  \\
\midrule

\multirow{21}{*}{0} 
& 0 & 1 & 1 & 0. & 0.292893 & 0 & 0.707107 & 0.0218184 & -3.52063 & 2.26795 & 0.94949 & 0 & 0 \\
 &1 & 0.999232 & 0.997291 & 0. & 0.267144 & 0.000768146 & 0.732087 & 0.021711 & -3.5383 & 2.27596 & 0.94949 & 0.0768146 & 0.0298324 \\
 &2 & 0.998464 & 0.994597 & 0. & 0.256703 & 0.00153635 & 0.741761 & 0.0216035 & -3.55614 & 2.28408 & 0.94949 & 0.0768173 & 0.0424519 \\
 &3 & 0.997695 & 0.991918 & 0. & 0.2488 & 0.0023046 & 0.748896 & 0.0214961 & -3.57414 & 2.29232 & 0.94949 & 0.0768199 & 0.0522663 \\
 &4 & 0.996927 & 0.989254 & 0. & 0.242218 & 0.0030729 & 0.754709 & 0.0213888 & -3.59233 & 2.30067 & 0.94949 & 0.0768225 & 0.060639 \\
 &5 & 0.996159 & 0.986605 & 0. & 0.236485 & 0.00384126 & 0.759674 & 0.0212815 & -3.61069 & 2.30914 & 0.94949 & 0.0768251 & 0.0680971 \\
& 6 & 0.99539 & 0.983971 & 0. & 0.231357 & 0.00460966 & 0.764033 & 0.0211741 & -3.62923 & 2.31774 & 0.94949 & 0.0768277 & 0.0749103 \\
& 7 & 0.994622 & 0.981351 & 0. & 0.226692 & 0.00537812 & 0.76793 & 0.0210668 & -3.64796 & 2.32647 & 0.94949 & 0.0768303 & 0.0812384 \\
 &8 & 0.993853 & 0.978747 & 0. & 0.222393 & 0.00614663 & 0.771461 & 0.0209595 & -3.66688 & 2.33532 & 0.94949 & 0.0768329 & 0.0871854 \\
 &9 & 0.993085 & 0.976158 & 0. & 0.218395 & 0.00691519 & 0.77469 & 0.0208522 & -3.686 & 2.34431 & 0.94949 & 0.0768354 & 0.0928234 \\
 &10 & 0.992316 & 0.973583 & 0. & 0.214651 & 0.0076838 & 0.777666 & 0.0207448 & -3.70532 & 2.35343 & 0.94949 & 0.076838 & 0.0982047 \\
 &11 & 0.991548 & 0.971024 & 0. & 0.211123 & 0.00845246 & 0.780424 & 0.0206374 & -3.72484 & 2.3627 & 0.94949 & 0.0768406 & 0.103369 \\
 &12 & 0.990779 & 0.968481 & 0. & 0.207785 & 0.00922117 & 0.782994 & 0.02053 & -3.74458 & 2.3721 & 0.94949 & 0.0768431 & 0.108346 \\
 &13 & 0.99001 & 0.965952 & 0. & 0.204613 & 0.00998993 & 0.785397 & 0.0204225 & -3.76453 & 2.38166 & 0.94949 & 0.0768456 & 0.113161 \\
& 14 & 0.989241 & 0.963439 & 0. & 0.201589 & 0.0107587 & 0.787652 & 0.0203149 & -3.78471 & 2.39137 & 0.94949 & 0.0768482 & 0.117834 \\
& 15 & 0.988472 & 0.960941 & 0. & 0.198698 & 0.0115276 & 0.789774 & 0.0202072 & -3.80512 & 2.40123 & 0.94949 & 0.0768507 & 0.12238 \\
& 16 & 0.987703 & 0.958459 & 0. & 0.195928 & 0.0122965 & 0.791776 & 0.0200994 & -3.82576 & 2.41126 & 0.94949 & 0.0768532 & 0.126813 \\
 &17 & 0.986935 & 0.955992 & 0. & 0.193267 & 0.0130655 & 0.793667 & 0.0199916 & -3.84665 & 2.42145 & 0.94949 & 0.0768557 & 0.131145 \\
& 18 & 0.986166 & 0.953541 & 0. & 0.190707 & 0.0138345 & 0.795459 & 0.0198836 & -3.86779 & 2.43182 & 0.94949 & 0.0768582 & 0.135385 \\
& 19 & 0.985396 & 0.951105 & 0. & 0.18824 & 0.0146035 & 0.797157 & 0.0197754 & -3.88919 & 2.44236 & 0.94949 & 0.0768607 & 0.139542 \\
& \textcolor{red}{20} & \textcolor{red}{0.984627} & \textcolor{red}{0.948685} & \textcolor{red}{0.} & \textcolor{red}{0.185858} & \textcolor{red}{0.0153726} & \textcolor{red}{0.798769} & \textcolor{red}{0.0196671} & \textcolor{red}{-3.91086} & \textcolor{red}{2.45308} & \textcolor{red}{0.94949} & \textcolor{red}{0.0768632} & \textcolor{red}{0.143622} \\
\midrule

\multirow{25}{*}{0.43} 
 &0 & 1 & 1 & 0.43 & 0.292893 & 0 & 0.707107 & 0.0221531 & -2.95759 & 2.29757 & 0.937916 & 0 & 0 \\
& 1 & 0.999345 & 0.997001 & 0.43 & 0.265867 & 0.000655197 & 0.733478 & 0.0220425 & -2.97267 & 2.30551 & 0.937916 & 0.0655197 & 0.0242429 \\
& 2 & 0.99869 & 0.994017 & 0.430282 & 0.254943 & 0.00131045 & 0.743746 & 0.0219326 & -2.98717 & 2.31358 & 0.937901 & 0.0655224 & 0.034531 \\
& 3 & 0.998034 & 0.991049 & 0.430847 & 0.246687 & 0.00196561 & 0.751347 & 0.0218233 & -3.00107 & 2.32178 & 0.93787 & 0.0655204 & 0.0425402 \\
 &4 & 0.997379 & 0.988096 & 0.431695 & 0.239818 & 0.00262055 & 0.757561 & 0.0217147 & -3.01437 & 2.33013 & 0.937823 & 0.0655136 & 0.0493744 \\
& 5 & 0.996725 & 0.985158 & 0.432829 & 0.233839 & 0.00327511 & 0.762886 & 0.0216068 & -3.02704 & 2.33861 & 0.93776 & 0.0655022 & 0.0554593 \\
& 6 & 0.996071 & 0.982233 & 0.434252 & 0.228494 & 0.00392915 & 0.767577 & 0.0214995 & -3.03905 & 2.34723 & 0.937681 & 0.0654859 & 0.0610125 \\
 &7 & 0.995417 & 0.979323 & 0.435965 & 0.223631 & 0.00458253 & 0.771786 & 0.021393 & -3.05038 & 2.35598 & 0.937586 & 0.0654648 & 0.0661622 \\
 &8 & 0.994765 & 0.976426 & 0.437972 & 0.219151 & 0.0052351 & 0.775614 & 0.0212872 & -3.06101 & 2.36487 & 0.937474 & 0.0654388 & 0.0709918 \\
 &9 & 0.994113 & 0.973543 & 0.440276 & 0.214984 & 0.0058867 & 0.779129 & 0.0211822 & -3.07088 & 2.37389 & 0.937344 & 0.0654078 & 0.0755586 \\
& 10 & 0.993463 & 0.970672 & 0.442884 & 0.21108 & 0.00653718 & 0.782383 & 0.021078 & -3.07998 & 2.38304 & 0.937197 & 0.0653718 & 0.079904 \\
& 11 & 0.992814 & 0.967813 & 0.445798 & 0.207401 & 0.00718638 & 0.785413 & 0.0209747 & -3.08825 & 2.39232 & 0.937031 & 0.0653308 & 0.0840589 \\
 &12 & 0.992166 & 0.964966 & 0.449025 & 0.203917 & 0.00783414 & 0.788249 & 0.0208723 & -3.09566 & 2.40174 & 0.936845 & 0.0652845 & 0.088047 \\
 &13 & 0.99152 & 0.96213 & 0.45257 & 0.200603 & 0.00848027 & 0.790917 & 0.0207709 & -3.10214 & 2.41127 & 0.93664 & 0.0652329 & 0.0918871 \\
 &14 & 0.990875 & 0.959305 & 0.456441 & 0.197441 & 0.00912461 & 0.793434 & 0.0206705 & -3.10766 & 2.42094 & 0.936414 & 0.0651758 & 0.0955939 \\
 &15 & 0.990233 & 0.956491 & 0.460644 & 0.194415 & 0.00976698 & 0.795818 & 0.0205712 & -3.11214 & 2.43072 & 0.936167 & 0.0651132 & 0.0991794 \\
& 16 & 0.989593 & 0.953687 & 0.465188 & 0.191511 & 0.0104072 & 0.798081 & 0.0204731 & -3.11552 & 2.44062 & 0.935896 & 0.0650449 & 0.102653 \\
& 17 & 0.988955 & 0.950891 & 0.47008 & 0.188718 & 0.011045 & 0.800237 & 0.0203762 & -3.11773 & 2.45063 & 0.935602 & 0.0649708 & 0.106024 \\
 &18 & 0.98832 & 0.948105 & 0.47533 & 0.186026 & 0.0116803 & 0.802293 & 0.0202806 & -3.11869 & 2.46075 & 0.935282 & 0.0648906 & 0.109298 \\
 &19 & 0.987687 & 0.945327 & 0.480947 & 0.183427 & 0.0123128 & 0.80426 & 0.0201864 & -3.11833 & 2.47097 & 0.934936 & 0.0648042 & 0.11248 \\
 &20 & 0.987058 & 0.942556 & 0.486943 & 0.180912 & 0.0129423 & 0.806146 & 0.0200938 & -3.11655 & 2.48129 & 0.934561 & 0.0647113 & 0.115576 \\
& 21 & 0.986431 & 0.939792 & 0.493328 & 0.178476 & 0.0135685 & 0.807955 & 0.0200028 & -3.11325 & 2.4917 & 0.934157 & 0.0646119 & 0.118588 \\
& 22 & 0.985809 & 0.937033 & 0.500114 & 0.176112 & 0.0141912 & 0.809696 & 0.0199136 & -3.10834 & 2.50219 & 0.933721 & 0.0645056 & 0.12152 \\
& 23 & 0.98519 & 0.934281 & 0.507313 & 0.173816 & 0.0148102 & 0.811374 & 0.0198263 & -3.1017 & 2.51275 & 0.933251 & 0.0643923 & 0.124375 \\
& \textcolor{red}{24} & \textcolor{red}{0.984575} & \textcolor{red}{0.931532} & \textcolor{red}{0.514939} & \textcolor{red}{0.171582} & \textcolor{red}{0.0154252} & \textcolor{red}{0.812993} & \textcolor{red}{0.0197411} & \textcolor{red}{-3.09321} & \textcolor{red}{2.52337} & \textcolor{red}{0.932745} & \textcolor{red}{0.0642716} & \textcolor{red}{0.127154} \\
\midrule

\multirow{24}{*}{0.55}
& 0 & 1 & 1 & 0.55 & 0.292893 & 0 & 0.707107 & 0.0223795 & -2.6309 & 2.31559 & 0.930313 & 0 & 0 \\
 &1 & 0.999411 & 0.996824 & 0.55 & 0.265122 & 0.000588781 & 0.734289 & 0.0222672 & -2.64441 & 2.32348 & 0.930313 & 0.0588781 & 0.0212009 \\
& 2 & 0.998822 & 0.993665 & 0.550324 & 0.253918 & 0.00117762 & 0.744905 & 0.0221558 & -2.65707 & 2.33152 & 0.930289 & 0.0588809 & 0.0302144 \\
& 3 & 0.998234 & 0.990521 & 0.550973 & 0.245457 & 0.00176631 & 0.752777 & 0.0220454 & -2.66885 & 2.3397 & 0.930243 & 0.0588771 & 0.0372356 \\
& 4 & 0.997645 & 0.987393 & 0.551948 & 0.238422 & 0.00235467 & 0.759223 & 0.0219359 & -2.67973 & 2.34803 & 0.930172 & 0.0588668 & 0.0432279 \\
& 5 & 0.997058 & 0.984279 & 0.55325 & 0.232301 & 0.00294249 & 0.764756 & 0.0218275 & -2.68967 & 2.3565 & 0.930078 & 0.0588499 & 0.0485625 \\
 &6 & 0.99647 & 0.98118 & 0.554883 & 0.226832 & 0.00352958 & 0.769639 & 0.02172 & -2.69866 & 2.36511 & 0.92996 & 0.0588263 & 0.0534287 \\
& 7 & 0.995884 & 0.978094 & 0.556849 & 0.221857 & 0.00411573 & 0.774028 & 0.0216137 & -2.70665 & 2.37386 & 0.929817 & 0.0587962 & 0.0579383 \\
& 8 & 0.995299 & 0.975022 & 0.55915 & 0.217274 & 0.00470074 & 0.778025 & 0.0215085 & -2.71362 & 2.38275 & 0.929649 & 0.0587592 & 0.0621633 \\
& 9 & 0.994716 & 0.971962 & 0.561791 & 0.213012 & 0.0052844 & 0.781703 & 0.0214044 & -2.71951 & 2.39177 & 0.929455 & 0.0587155 & 0.0661536 \\
& 10 & 0.994134 & 0.968914 & 0.564775 & 0.20902 & 0.00586649 & 0.785114 & 0.0213016 & -2.72428 & 2.40093 & 0.929235 & 0.0586649 & 0.0699447 \\
& 11 & 0.993553 & 0.965877 & 0.568108 & 0.205257 & 0.00644681 & 0.788296 & 0.0212 & -2.7279 & 2.41021 & 0.928987 & 0.0586073 & 0.0735634 \\
& 12 & 0.992975 & 0.962851 & 0.571794 & 0.201693 & 0.00702512 & 0.791282 & 0.0210998 & -2.73029 & 2.41962 & 0.928711 & 0.0585427 & 0.0770299 \\
& 13 & 0.992399 & 0.959834 & 0.57584 & 0.198304 & 0.00760121 & 0.794095 & 0.021001 & -2.73141 & 2.42915 & 0.928406 & 0.0584708 & 0.0803601 \\
& 14 & 0.991825 & 0.956828 & 0.58025 & 0.195069 & 0.00817483 & 0.796756 & 0.0209036 & -2.73119 & 2.4388 & 0.928071 & 0.0583916 & 0.0835666 \\
& 15 & 0.991254 & 0.953829 & 0.585033 & 0.191972 & 0.00874574 & 0.799282 & 0.0208078 & -2.72956 & 2.44856 & 0.927705 & 0.058305 & 0.0866594 \\
 &16 & 0.990686 & 0.950839 & 0.590195 & 0.189 & 0.00931371 & 0.801686 & 0.0207137 & -2.72647 & 2.45842 & 0.927305 & 0.0582107 & 0.0896468 \\
 &17 & 0.990122 & 0.947856 & 0.595743 & 0.18614 & 0.00987846 & 0.803982 & 0.0206214 & -2.72182 & 2.46839 & 0.926872 & 0.0581086 & 0.0925353 \\
 &18 & 0.98956 & 0.94488 & 0.601687 & 0.183382 & 0.0104397 & 0.806178 & 0.0205308 & -2.71554 & 2.47846 & 0.926402 & 0.0579985 & 0.0953304 \\
 &19 & 0.989003 & 0.941909 & 0.608035 & 0.180718 & 0.0109973 & 0.808285 & 0.0204423 & -2.70755 & 2.48861 & 0.925895 & 0.0578803 & 0.0980364 \\
 &20 & 0.988449 & 0.938942 & 0.614796 & 0.17814 & 0.0115507 & 0.81031 & 0.0203559 & -2.69775 & 2.49885 & 0.925348 & 0.0577537 & 0.100657 \\
& 21 & 0.9879 & 0.93598 & 0.62198 & 0.17564 & 0.0120999 & 0.81226 & 0.0202717 & -2.68605 & 2.50916 & 0.924759 & 0.0576186 & 0.103195 \\
& 22 & 0.987356 & 0.93302 & 0.629598 & 0.173214 & 0.0126444 & 0.814141 & 0.0201898 & -2.67234 & 2.51953 & 0.924127 & 0.0574746 & 0.105653 \\
& \textcolor{blue}{23} & \textcolor{blue}{0.986816} & \textcolor{blue}{0.930063} & \textcolor{blue}{0.637661} & \textcolor{blue}{0.170855} & \textcolor{blue}{0.0131839} & \textcolor{blue}{0.815961} & \textcolor{blue}{0.0201105} & \textcolor{blue}{-2.65652} & \textcolor{blue}{2.52996} & \textcolor{blue}{0.923448} & \textcolor{blue}{0.0573215} & \textcolor{blue}{0.108032} \\
\midrule

\multirow{3}{*}{0.88}
& 0 & 1 & 1 & 0.88 & 0.292893 & 0 & 0.707107 & 0.0233651 & -1.5245 & 2.38245 & 0.898404 & 0 & 0 \\
& 1 & 0.999644 & 0.996169 & 0.88 & 0.262529 & 0.0003562 & 0.737115 & 0.0232467 & -1.53251 & 2.39019 & 0.898404 & 0.03562 & 0.0117308 \\
& \textcolor{blue}{2} & \textcolor{blue}{0.999288} & \textcolor{blue}{0.992357} & \textcolor{blue}{0.880314} & \textcolor{blue}{0.250354} & \textcolor{blue}{0.000712458} & \textcolor{blue}{0.748934} & \textcolor{blue}{0.0231299} & \textcolor{blue}{-1.53944} & \textcolor{blue}{2.39809} & \textcolor{blue}{0.898365} & \textcolor{blue}{0.0356229} & \textcolor{blue}{0.0167482} \\
\midrule

\multirow{2}{*}{0.92}
& 0 & 1 & 1 & 0.92 & 0.292893 & 0 & 0.707107 & 0.0235237 & -1.38033 & 2.39198 & 0.893384 & 0 & 0 \\
& \textcolor{blue}{1} & \textcolor{blue}{0.999675} & \textcolor{blue}{0.996075} & \textcolor{blue}{0.92} & \textcolor{blue}{0.262178} & \textcolor{blue}{0.000324703} & \textcolor{blue}{0.737497} & \textcolor{blue}{0.0234045} & \textcolor{blue}{-1.38761} & \textcolor{blue}{2.39971} & \textcolor{blue}{0.893384} & \textcolor{blue}{0.0324703} & \textcolor{blue}{0.0105714} \\
\bottomrule
\end{tabular}%
}%
\caption{RPP for neutral particles at $\hat{r}=1.5$ for various initial $\xi_0$ in rotating R-S-V Black Hole. The red and blue rows shown the kinematic spin threshold and structural termination condition respectively. The physical RPP ends at the preceding row.}
\label{2}
\end{table*}
As the regularization parameter increases, the iterative evolution exhibits a distinctly different behaviour. Once $\hat{\xi}$ exceeds the above threshold, before the minimum-spin condition is attained, the system reaches the boundary of the parameter domain defining the initial spacetime geometry. At this stage, the evolving solution no longer belongs to the two-horizon branch on which the iterative framework is constructed. This behaviour is illustrated by the case $\hat{\xi}=0.55$. Although the black hole still possesses sufficient rotational energy to support additional Penrose events according to the conventional spin criterion, the evolving configuration has already crossed the geometrical boundary associated with the initial branch. The iterative sequence therefore terminates by the geometry-induced constraint rather than by the exhaustion of rotational energy. The same tendency becomes progressively stronger as $\hat{\xi}$ approaches the upper limit of the two-horizon regime. For example, the results obtained for $\hat{\xi}=0.88$ show a substantial reduction in the number of admissible Penrose iterations, while the dynamical spin limit no longer determines the endpoint of the process. Increasing the regularization parameter therefore changes not only the amount of rotational energy that can be extracted through the RPP, but also the mechanism that ultimately terminates the iterative sequence.

An even more restrictive regime is reached near $\hat{\xi}\approx0.9$. In this regime, a single Penrose event is sufficient to move the evolving solution outside the admissible parameter domain of the initial two-horizon configuration. Consequently, the iterative sequence terminates after only a single Penrose event, despite the fact that the black hole has not yet reached its minimum spin. Accordingly, the geometry-induced termination is best interpreted as a limitation of the physical regime under consideration rather than as a universal prohibition of rotational energy extraction. The present calculations do not imply that subsequent Penrose processes become impossible in general once the geometrical boundary is crossed. Rather, they indicate that the assumptions defining the present iterative framework are no longer satisfied. Since the evolving solution no longer belongs to the same branch of the R-S-V spacetime, any further analysis would require reformulating the RPP within the geometry corresponding to the new branch. The endpoint identified here therefore marks the limit of validity of the original two-horizon black-hole description, not necessarily the end of the physical evolution of the spacetime itself.

The existence of a geometry-induced termination also raises a broader question. If an astrophysical compact object experiences a succession of physical processes capable of continuously modifying its mass and angular momentum over sufficiently long timescales, could its trajectory in parameter space eventually intersect different branches of the S-V solution family and push the model towards becoming a wormhole? The present analysis neither establishes nor rules out such a possibility. Instead, the geometry-induced termination identified here provides a natural starting point for investigating this problem within more complete dynamical frameworks. More broadly, the present results suggest that the RPP may also provide a useful probe of how the global properties of a rotating spacetime evolve under successive energy-extraction events. 
\subsection{Analysis of RPP energy extraction considering particle release distance} 
In this section, we investigate how the spacetime regularization parameter influences the cumulative outcome of the RPP. To this end, we examine the final extracted energy, the final irreducible mass (both normalized to the initial black-hole mass), together with the Energy Return on Investment (EROI) and the Energy Utilization Efficiency (EUE), as functions of the dimensionless decay radius. Considered together, these quantities provide complementary information on both the dynamical evolution of the black hole and the thermodynamic efficiency of the repetitive extraction process.
\begin{figure}[htbp]
    \centering
    \includegraphics[width=1.01\linewidth]{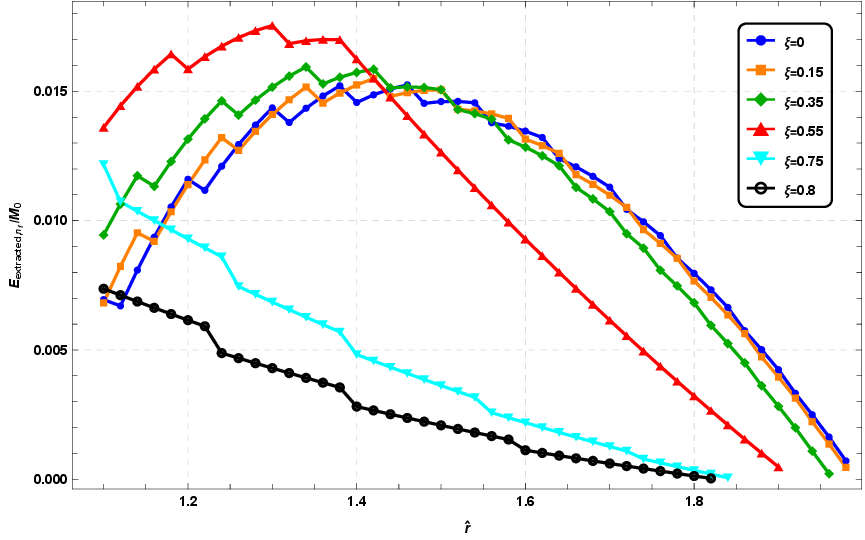}
    \caption{ The behavior of Extracted energy, normalized to the initial black hole mass, as a function of the dimensionless decay radius $\hat{r}$ for various values of the deformation parameter $\xi$.}
    \label{2}
\end{figure}
Figure~\ref{2} shows the cumulative extracted energy at the termination of the RPP. Two qualitatively distinct regimes can be identified, depending on the value of the regularization parameter $\xi$.

For relatively small values of $\xi$, the overall behavior remains broadly similar to that of the Kerr spacetime. Nevertheless, the regularization modifies both the magnitude and the radial location of the optimal extraction region. As $\xi$ increases from zero, the extracted energy develops a pronounced maximum at an intermediate decay radius, with the largest enhancement occurring around $\xi=0.55$. At the same time, the position of this maximum shifts progressively toward smaller radii, indicating that the decay radius yielding the highest cumulative energy extraction moves closer to the event horizon as the regularization becomes stronger.

A qualitatively different behavior appears for larger values of the regularization parameter ($\xi=0.75$ and $0.80$). In this regime, the extracted energy is substantially reduced over the entire admissible radial range. The internal maximum disappears, the largest extracted energy is obtained at the smallest allowed decay radius, and the curves decrease monotonically with increasing radius. This suppression reflects not only the reduced amount of rotational energy that can be extracted during the repetitive process but also the fact that the geometry-induced termination discussed in the previous section substantially shortens the sequence of admissible Penrose events. Consequently, the available parameter space for sustained repetitive extraction becomes progressively more restricted as $\xi$ approaches the upper boundary of the two-horizon regime. 
 
 \begin{figure}[htbp]
    \centering
    \includegraphics[width=1.01\linewidth]{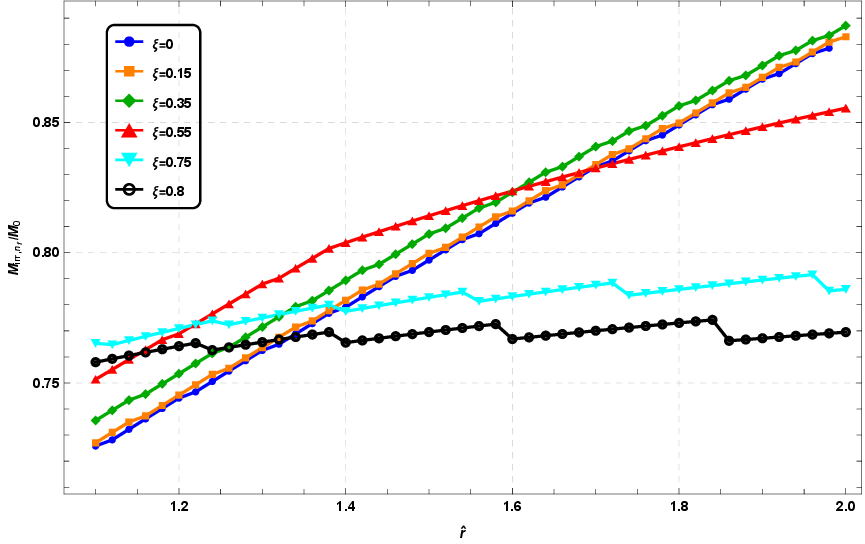}
    \caption{ The behavior of the final irreducible mass, normalized to the initial black hole mass ($M_{\text{irr}, n_f}/M_0$), as a function of the dimensionless decay radius $\hat{r}$ for various values of the deformation parameter $\xi$. }
    \label{3}
\end{figure} 
Figure~\ref{3} presents the final irreducible mass reached at the termination of the RPP. As in the extracted-energy analysis, two distinct regimes emerge.

For relatively small values of the regularization parameter ($\xi\le0.55$), the irreducible mass increases monotonically with the decay radius, consistent with the irreversible growth of the event-horizon area required by Hawking's area theorem. The dependence on $\xi$, however, is not uniform throughout the radial domain. Near the horizon ($\hat r\lesssim1.4$), increasing $\xi$ produces a larger final irreducible mass, implying that a greater fraction of the initial rotational-energy reservoir is ultimately converted into horizon area rather than escaping energy. At larger radii, the ordering of the curves changes, producing crossover points that indicate a redistribution of the relative thermodynamic efficiency among different values of the regularization parameter.

For larger values of $\xi$ ($\xi\ge0.75$), the behavior changes markedly. The growth of the irreducible mass becomes much weaker, and the curves evolve toward an almost flat profile with visible slope discontinuities. These features should not be interpreted as a breakdown of the thermodynamic behavior of the spacetime. Instead, they primarily reflect the premature termination of the repetitive Penrose sequence caused by the geometry-induced constraint. Since the iterative evolution ends after a much smaller number of admissible decay events, the cumulative increase of the irreducible mass is likewise strongly limited.
 \begin{figure}[htbp]
    \centering
    \includegraphics[width=1.01\linewidth]{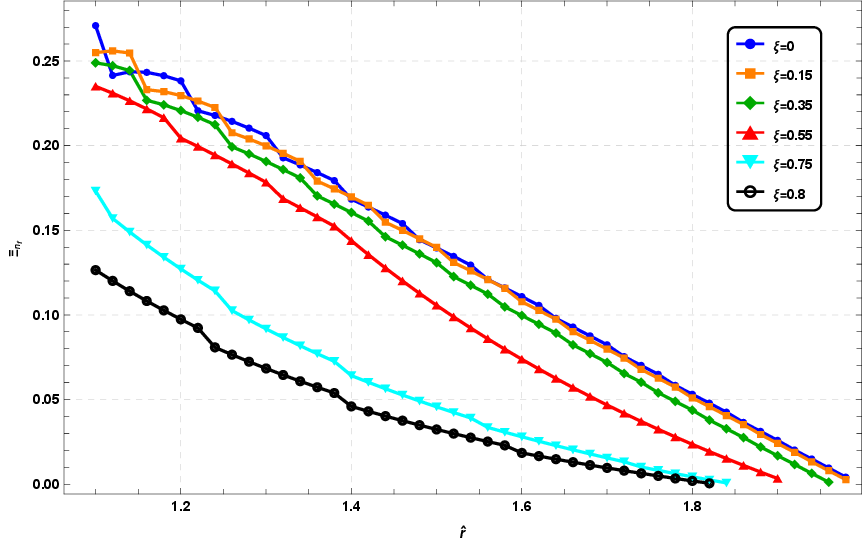}
    \caption{ The final energy utilization efficiency $\Xi_{nf}$ as a function of the dimensionless decay radius $\hat{r}$ for various values of the deformation parameter $\xi$. }
    \label{4}
\end{figure} 
Figure~\ref{4} presents the final Energy Utilization Efficiency (EUE) as a function of the dimensionless decay radius. For all values of the regularization parameter, the efficiency decreases monotonically as the decay radius increases, indicating that repetitive Penrose events occurring closer to the event horizon convert a larger fraction of the available extractable rotational energy into escaping energy.

For relatively small values of the regularization parameter ($\xi\lesssim0.55$), the efficiency curves remain smooth and retain a behavior qualitatively similar to that of the Kerr spacetime, although the overall efficiency gradually decreases as $\xi$ increases.

A markedly different regime appears for larger values of the regularization parameter ($\xi\gtrsim0.75$). In this case, the EUE is substantially reduced over the entire admissible radial interval. This reduction should not be interpreted solely as a consequence of the local properties of the spacetime. Instead, it primarily reflects the cumulative effect of the geometry-induced termination discussed in the previous section. Since the repetitive Penrose sequence ends after considerably fewer admissible decay events, only a smaller fraction of the initially available rotational-energy reservoir can ultimately be converted into escaping energy. Consequently, the EUE provides an integrated measure of how strongly spacetime regularization limits the long-term thermodynamic performance of the repetitive extraction process.
\begin{figure}[htbp]
    \centering
    \includegraphics[width=1.01\linewidth]{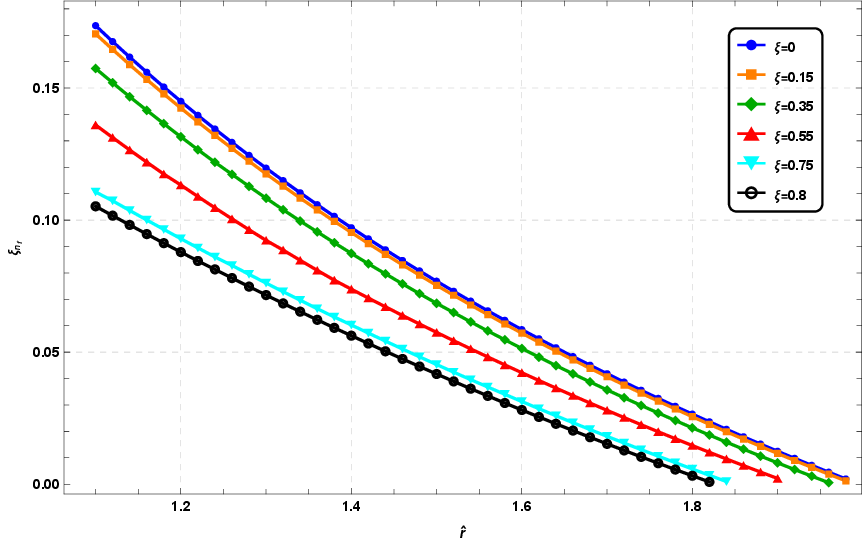}
    \caption{ The behavior of the final energy return on investment (EROI), denoted as $\xi_{n_f}$, as a function of the dimensionless decay radius $\hat{r}$ for various values of the regularization parameter $\xi$. }
    \label{5}
\end{figure} 

Figure~\ref{5} shows the final Energy Return on Investment (EROI) of the RPP. All curves decrease monotonically with increasing decay radius, indicating that energy extraction becomes progressively less advantageous as the decay point moves away from the near-horizon region.

For every fixed decay radius, increasing the regularization parameter systematically lowers the EROI. The Kerr spacetime ($\xi=0$) consistently provides the largest energy return, while stronger regularization shifts the curves downward over the entire admissible parameter range. This suppression is particularly pronounced near the event horizon, where the cumulative effects of repetitive extraction are otherwise expected to be most efficient.

Unlike the extracted energy alone, the EROI simultaneously reflects both the cumulative energy gained from the RPP and the total energy invested through the injected particles. Its systematic reduction therefore indicates not only that less rotational energy is ultimately extracted, but also that the repetitive sequence becomes progressively shorter as the geometry-induced termination sets in. Accordingly, the deterioration of the EROI quantifies the combined dynamical and geometrical limitations imposed by increasing spacetime regularization.

Taken together, Figs. 2-5 reveal a consistent physical picture. Increasing the regularization parameter influences the RPP through two complementary mechanisms. First, it reduces the cumulative amount of rotational energy that can be extracted. Second, and more importantly, it progressively shortens the admissible repetitive sequence through the geometry-induced termination identified in the previous section. These two effects jointly suppress the extracted energy, the growth of the irreducible mass, the energy utilization efficiency, and the energy return on investment, demonstrating that spacetime regularization constrains not only the energetics of the process but also its long-term dynamical evolution.
\section{Conclusion}
In this work, we have investigated the RPP in the R-S-V spacetime to examine how the underlying geometry influences the cumulative extraction of rotational energy. Unlike the Kerr solution, the R-S-V geometry admits several geometrically distinct branches within a single family of solutions, characterized by the regularization parameter $\xi$. This additional degree of freedom provides a natural framework for exploring whether the long-term evolution of repetitive energy extraction is constrained not only by the dynamics of the Penrose process itself but also by the geometrical structure of the background spacetime.

Assuming that the regularization parameter remains fixed throughout the evolution, we followed the successive changes in the black-hole mass and angular momentum produced by repeated Penrose events. Our analysis shows that, for sufficiently large values of the regularization parameter, the endpoint of the repetitive process is no longer determined solely by the conventional minimum-spin condition. Instead, the evolving solution may leave the parameter domain associated with the initial two-horizon black-hole branch before the kinematic spin threshold is reached. Within the framework considered here, this introduces an additional geometry-induced termination of the repetitive sequence.

The numerical results further show that increasing the regularization parameter progressively limits both the cumulative amount of rotational energy that can be extracted and the duration of the repetitive Penrose sequence. This combined effect is reflected in the reduction of the total extracted energy, the evolution of the irreducible mass, and the systematic suppression of both the EUE and the EROI. In addition, the particle decay radius plays a significant role in determining the overall performance of the process, with near-horizon decays consistently providing the most favorable conditions for cumulative energy extraction.

A central outcome of the present analysis is the identification of two conceptually distinct mechanisms governing the termination of the RPP. The first is the familiar dynamical limitation associated with the minimum spin required to sustain negative-energy trajectories. The second arises from the geometrical restrictions imposed by the admissible parameter space of the R-S-V solution. The competition between these two mechanisms provides a unified physical picture of how modifications of the spacetime geometry can influence the cumulative evolution of repetitive energy extraction.

Finally, we emphasize the scope of the present results. The geometry-induced termination identified here should not be interpreted as implying that rotational energy extraction becomes fundamentally impossible once the corresponding parameter boundary is crossed. Rather, it indicates that the assumptions underlying the present iterative framework are no longer satisfied. Describing the subsequent evolution would require reformulating the RPP within the spacetime branch corresponding to the evolved black-hole parameters, which lies beyond the scope of the present work.

More broadly, the present study shows that the RPP provides not only an idealized framework for investigating rotational energy extraction but also a useful tool for exploring how the global evolution of a rotating spacetime is constrained by its underlying geometry. Future investigations extending the present framework to more general regular rotating spacetimes, charged configurations, or models in which the geometric parameters evolve dynamically may clarify whether geometry-induced termination represents a generic feature of repetitive energy extraction beyond the R-S-V spacetime.
\section{Appendix}
\textbf{ $\mathcal{A}$, $\mathcal{B}$ and  $\mathcal{D}$ Definitions in equation 9}:\\
$\mathcal{A}$, $\mathcal{B}$, and the discriminant $\mathcal{D}$ are purely algebraic functions of the metric components evaluated at the decay radius, the black hole parameters, and the initial kinematic variables:
\begin{equation*}
\mathcal{A}=\hat E_0\hat E_1 g^{tt} - \hat E_1 g^{t\phi}M\hat p_{\phi 0} - \hat E_0 g^{t\phi}M\hat p_{\phi 1}+ g^{\phi\phi}M^2\hat p_{\phi 0}\hat p_{\phi 1}, 
\end{equation*}
\begin{equation*}
\mathcal{B}=\hat E_1^2 g^{tt} - 2\hat E_1 g^{t\phi}M\hat p_{\phi 1}+ g^{\phi\phi}M^2\hat p_{\phi 1}^2 + \nu^2,
\end{equation*}
\begin{equation*}
\begin{split}
&\mathcal{D}=\\& -g^{tt}g^{\phi\phi}M^2\hat E_1^2\hat p_{\phi 0}^2+ (g^{t\phi})^2M^2\hat E_1^2\hat p_{\phi 0}^2
- g^{tt}g^{\phi\phi}M^2\hat E_0^2\hat p_{\phi 1}^2\\&+(g^{t\phi})^2M^2\hat E_0^2\hat p_{\phi 1}^2 + 2g^{tt}g^{\phi\phi}M^2\hat E_0\hat E_1\hat p_{\phi 0}\hat p_{\phi 1}-\\& 2(g^{t\phi})^2M^2\hat E_0\hat E_1\hat p_{\phi 0}\hat p_{\phi 1}- g^{tt}\hat E_0^2\nu^2 \\&+ 2g^{t\phi}M\hat E_0\hat p_{\phi 0}\nu^2 
- g^{\phi\phi}M^2\hat p_{\phi 0}^2\nu^2 .
\end{split}
\end{equation*}\\
\textbf{The repetitive Penrose process at $\hat{r}=1.2$}\\
A comparison of Tables 2 and 3 clearly demonstrates the competition between the Dynamical and Geometry Induced Termination conditions. In particular, for the choice of $\xi=0.55$, it can be clearly seen that for $\hat{r}=1.2$ it is the dynamical condition that terminates the process, and for 1.5 it is the geometry that decides. 
\begin{table*}[p]
\centering
\setlength{\arrayrulewidth}{0.5pt}
\renewcommand{\arraystretch}{1.2}
\scriptsize
\setlength{\tabcolsep}{3pt}
\resizebox{\textwidth}{!}{%
\begin{tabular}{cccccccccccccc}
\toprule
 $\xi_0$ & $n$ & $M/M_0$ & $\hat{a}_n$ & $\hat{\xi}_n$ & $E_{\text{extractable}}/M_0$ & $E_{\text{extracted}}/M_0$ & $M_\text{irr}/M_0$ & $\tilde{\mu}_{1,n}$ & $\hat{E}_{1,n}$ & $\hat{p}_{\phi0,n}$ & $\hat{a}_{\min,0,n}$ & $\zeta_n$ & $\Xi_n$  \\
\midrule

\multirow{10}{*}{0} 
 & 0 & 1 & 1 & 0 & 0.292893 & 0 & 0.707107 & 0.0209006 & -6.93558 & 2.1127 & 0.99089 & 0 & 0 \\
 & 1 & 0.99855 & 0.998832 & 0 & 0.275611 & 0.00144958 & 0.72294 & 0.0208126 & -6.96513 & 2.12126 & 0.99089 & 0.144958 & 0.0838746 \\
 & 2 & 0.997101 & 0.997676 & 0 & 0.268419 & 0.0028992 & 0.728682 & 0.0207235 & -6.99529 & 2.13006 & 0.99089 & 0.14496 & 0.11846 \\
 & 3 & 0.995651 & 0.996532 & 0 & 0.262915 & 0.00434887 & 0.732736 & 0.0206333 & -7.0261 & 2.13914 & 0.99089 & 0.144962 & 0.145066 \\
 & 4 & 0.994201 & 0.995402 & 0 & 0.258294 & 0.00579858 & 0.735907 & 0.0205417 & -7.05764 & 2.14852 & 0.99089 & 0.144965 & 0.167593 \\
 & 5 & 0.992752 & 0.994284 & 0 & 0.254245 & 0.00724834 & 0.738506 & 0.0204488 & -7.08996 & 2.15822 & 0.99089 & 0.144967 & 0.187549 \\
 & 6 & 0.991302 & 0.99318 & 0 & 0.250608 & 0.00869815 & 0.740694 & 0.0203542 & -7.12314 & 2.16827 & 0.99089 & 0.144969 & 0.205702 \\
 & 7 & 0.989852 & 0.992089 & 0 & 0.247286 & 0.010148 & 0.742566 & 0.0202578 & -7.15727 & 2.17872 & 0.99089 & 0.144971 & 0.22251 \\
 & 8 & 0.988402 & 0.991013 & 0 & 0.244218 & 0.0115979 & 0.744184 & 0.0201593 & -7.19248 & 2.18961 & 0.99089 & 0.144974 & 0.238271 \\
 & \textcolor{red}{9} & \textcolor{red}{0.986952} & \textcolor{red}{0.989952} & \textcolor{red}{0} & \textcolor{red}{0.24136} & \textcolor{red}{0.0130479} & \textcolor{red}{0.745592} & \textcolor{red}{0.0200585} & \textcolor{red}{-7.22889} & \textcolor{red}{2.20099} & \textcolor{red}{0.99089} & \textcolor{red}{0.144976} & \textcolor{red}{0.253193} \\
\midrule

\multirow{13}{*}{0.43} 
 & 0 & 1 & 1 & 0.43 & 0.292893 & 0 & 0.707107 & 0.0209941 & -5.99043 & 2.15276 & 0.98335 & 0 & 0 \\
 & 1 & 0.998742 & 0.99843 & 0.43 & 0.273014 & 0.00125763 & 0.725728 & 0.0209004 & -6.01749 & 2.16114 & 0.98335 & 0.125763 & 0.0632635 \\
 & 2 & 0.997485 & 0.996872 & 0.430541 & 0.264814 & 0.00251532 & 0.732671 & 0.0208067 & -6.04267 & 2.16979 & 0.98333 & 0.125766 & 0.0895793 \\
  &3 & 0.996227 & 0.995326 & 0.431627 & 0.258556 & 0.0037726 & 0.737671 & 0.0207128 & -6.06593 & 2.17871 & 0.983288 & 0.125753 & 0.10987 \\
 & 4 & 0.994971 & 0.99379 & 0.433262 & 0.25331 & 0.00502902 & 0.741661 & 0.020619 & -6.08718 & 2.1879 & 0.983224 & 0.125726 & 0.127051 \\
  &5 & 0.993716 & 0.992265 & 0.435452 & 0.248714 & 0.00628413 & 0.745002 & 0.0205252 & -6.10632 & 2.19735 & 0.983139 & 0.125683 & 0.142241 \\
 & 6 & 0.992463 & 0.990748 & 0.438205 & 0.24458 & 0.00753747 & 0.747883 & 0.0204315 & -6.12325 & 2.20705 & 0.98303 & 0.125624 & 0.156013 \\
  &7 & 0.991211 & 0.98924 & 0.441533 & 0.240797 & 0.00878854 & 0.750414 & 0.0203381 & -6.13782 & 2.217 & 0.982898 & 0.125551 & 0.168699 \\
  &8 & 0.989963 & 0.987739 & 0.445448 & 0.237293 & 0.0100369 & 0.75267 & 0.0202452 & -6.14987 & 2.22719 & 0.982741 & 0.125461 & 0.180517 \\
  &9 & 0.988718 & 0.986243 & 0.449964 & 0.234015 & 0.0112819 & 0.754703 & 0.0201528 & -6.15919 & 2.2376 & 0.982557 & 0.125355 & 0.191614 \\
  &10 & 0.987477 & 0.984753 & 0.455099 & 0.230926 & 0.0125232 & 0.756551 & 0.0200612 & -6.16556 & 2.24821 & 0.982345 & 0.125232 & 0.202094 \\
  &11 & 0.98624 & 0.983267 & 0.46087 & 0.227998 & 0.01376 & 0.758242 & 0.0199706 & -6.16873 & 2.259 & 0.982103 & 0.125091 & 0.212035 \\
& \textcolor{red}{12} & \textcolor{red}{0.985008} & \textcolor{red}{0.981783} & \textcolor{red}{0.4673} & \textcolor{red}{0.225208} & \textcolor{red}{0.014992} & \textcolor{red}{0.7598} & \textcolor{red}{0.0198813} & \textcolor{red}{-6.1684} & \textcolor{red}{2.26994} & \textcolor{red}{0.981828} & \textcolor{red}{0.124933} & \textcolor{red}{0.221495} \\
\midrule

\multirow{16}{*}{0.55}
 &0 & 1 & 1 & 0.55 & 0.292893 & 0 & 0.707107 & 0.021099 & -5.44673 & 2.17658 & 0.97782 & 0 & 0 \\
 &1 & 0.998851 & 0.998193 & 0.55 & 0.271643 & 0.0011492 & 0.727208 & 0.0210022 & -5.47205 & 2.18487 & 0.97782 & 0.11492 & 0.0540803 \\
 &2 & 0.997702 & 0.996398 & 0.550633 & 0.262916 & 0.00229846 & 0.734786 & 0.0209059 & -5.49456 & 2.19345 & 0.977786 & 0.114923 & 0.0766723 \\
 &3 & 0.996553 & 0.994615 & 0.551901 & 0.256267 & 0.00344714 & 0.740286 & 0.0208101 & -5.51416 & 2.2023 & 0.977717 & 0.114905 & 0.094116 \\
 &4 & 0.995405 & 0.992841 & 0.55381 & 0.250699 & 0.00459464 & 0.744707 & 0.0207149 & -5.53076 & 2.21143 & 0.977613 & 0.114866 & 0.108892 \\
 &5 & 0.99426 & 0.991077 & 0.556367 & 0.245823 & 0.00574033 & 0.748437 & 0.0206204 & -5.54422 & 2.22081 & 0.977473 & 0.114807 & 0.121952 \\
 &6 & 0.993116 & 0.989321 & 0.559579 & 0.24144 & 0.00688357 & 0.751677 & 0.0205268 & -5.5544 & 2.23045 & 0.977296 & 0.114726 & 0.133782 \\
& 7 & 0.991976 & 0.98757 & 0.563457 & 0.237429 & 0.00802371 & 0.754548 & 0.0204343 & -5.56113 & 2.24031 & 0.977081 & 0.114624 & 0.144664 \\
 &8 & 0.99084 & 0.985824 & 0.568015 & 0.233712 & 0.00916009 & 0.757128 & 0.0203431 & -5.5642 & 2.25039 & 0.976825 & 0.114501 & 0.154779 \\
 &9 & 0.989708 & 0.984081 & 0.573266 & 0.230234 & 0.010292 & 0.759474 & 0.0202533 & -5.56339 & 2.26067 & 0.976526 & 0.114356 & 0.164253 \\
 &10 & 0.988581 & 0.98234 & 0.579228 & 0.226955 & 0.0114188 & 0.761627 & 0.0201653 & -5.55844 & 2.27112 & 0.976182 & 0.114188 & 0.173173 \\
 &11 & 0.98746 & 0.980598 & 0.585918 & 0.223843 & 0.0125397 & 0.763617 & 0.0200793 & -5.54908 & 2.28172 & 0.97579 & 0.113997 & 0.181603 \\
 &12 & 0.986346 & 0.978854 & 0.593359 & 0.220875 & 0.0136539 & 0.765471 & 0.0199957 & -5.53499 & 2.29245 & 0.975346 & 0.113782 & 0.18959 \\
& 13 & 0.985239 & 0.977106 & 0.601572 & 0.218031 & 0.0147606 & 0.767208 & 0.0199147 & -5.51583 & 2.30326 & 0.974847 & 0.113543 & 0.197171 \\
 &14 & 0.984141 & 0.975351 & 0.610585 & 0.215295 & 0.0158591 & 0.768846 & 0.0198369 & -5.49124 & 2.31414 & 0.974288 & 0.113279 & 0.204374 \\
 & \textcolor{red}{15} & \textcolor{red}{0.983052} & \textcolor{red}{0.973588} & \textcolor{red}{0.620424} & \textcolor{red}{0.212653} & \textcolor{red}{0.0169484} & \textcolor{red}{0.770399} & \textcolor{red}{0.0197625} & \textcolor{red}{-5.46081} & \textcolor{red}{2.32504} & \textcolor{red}{0.973665} & \textcolor{red}{0.112989} & \textcolor{red}{0.21122} \\
\midrule

\multirow{4}{*}{0.88}
 &0 & 1 & 1 & 0.88 & 0.292893 & 0 & 0.707107 & 0.0217575 & -3.63656 & 2.26205 & 0.951657 & 0 & 0 \\
& 1 & 0.999209 & 0.997349 & 0.88 & 0.267408 & 0.000791225 & 0.731801 & 0.0216507 & -3.65475 & 2.27007 & 0.951657 & 0.0791225 & 0.031046 \\
 &2 & 0.998417 & 0.994713 & 0.880697 & 0.257066 & 0.0015825 & 0.741352 & 0.021546 & -3.66901 & 2.2784 & 0.951582 & 0.0791251 & 0.04417 \\
 & \textcolor{blue}{3} & \textcolor{blue}{0.997627} & \textcolor{blue}{0.99209} & \textcolor{blue}{0.882093} & \textcolor{blue}{0.24923} & \textcolor{blue}{0.00237303} & \textcolor{blue}{0.748397} & \textcolor{blue}{0.0214438} & \textcolor{blue}{-3.67925} & \textcolor{blue}{2.28703} & \textcolor{blue}{0.951433} & \textcolor{blue}{0.0791009} & \textcolor{blue}{0.0543486} \\

\midrule
\multirow{3}{*}{0.92}
 & 0 & 1 & 1 & 0.92 & 0.292893 & 0 & 0.707107 & 0.021882 & -3.4047 & 2.27391 & 0.947252 & 0 & 0 \\
&1 & 0.999255 & 0.997233 & 0.92 & 0.266882 & 0.000745016 & 0.732373 & 0.0217739 & -3.42185 & 2.28191 & 0.947252 & 0.0745016 & 0.0286416 \\
& \textcolor{blue}{2} & \textcolor{blue}{0.99851} & \textcolor{blue}{0.994481} & \textcolor{blue}{0.920686} & \textcolor{blue}{0.256341} & \textcolor{blue}{0.00149008} & \textcolor{blue}{0.742169} & \textcolor{blue}{0.0216682} & \textcolor{blue}{-3.43511} & \textcolor{blue}{2.2902} & \textcolor{blue}{0.947174} & \textcolor{blue}{0.0745042} & \textcolor{blue}{0.0407655} \\

\bottomrule
\end{tabular}%
}%
\caption{RPP for neutral particles at $\hat{r}=1.2$ for various initial $\xi_0$ in rotating R-S-V Black Hole. The red and blue rows shown the kinematic spin threshold and structural termination condition respectively. The physical RPP ends at the preceding row.}
\label{3}
\end{table*}

\end{document}